\documentclass[aps,prc,superscriptaddress,twocolumn]{revtex4}
\usepackage{graphicx}% Include figure files
\usepackage{epsfig}%
\usepackage{bm}% bold math
\usepackage{hyperref}% add hypertext capabilities
\usepackage{amsmath} \usepackage{amssymb}
\usepackage{color}
\begin{document}

	\title{Bayesian reconstruction of impact parameter distributions from two observables for intermediate energy heavy ion collisions}

 	\author{Xiang Chen}
	\affiliation{China Institute of Atomic Energy, P. O. Box 275(18), Beijing 102413, China}
	\author{Li Li\footnote{The author contribute  equally to this paper}}
	\affiliation{China Institute of Atomic Energy, P. O. Box 275(18), Beijing 102413, China}
	\author{Ying Cui}
        \email{cuiying@ciae.ac.cn}
	\affiliation{China Institute of Atomic Energy, P. O. Box 275(18), Beijing 102413, China}
 	\author{Junping Yang}
	\affiliation{China Institute of Atomic Energy, P. O. Box 275(18), Beijing 102413, China}
	\author{Zhuxia Li}
	\affiliation{China Institute of Atomic Energy, P. O. Box 275(18), Beijing 102413, China}
	\author{Yingxun Zhang}
	\email{zhyx@ciae.ac.cn}
	\affiliation{China Institute of Atomic Energy, P. O. Box 275(18), Beijing 102413, China}%
	\affiliation{Department of Physics and Technology, Guangxi normal University, Guilin 540101, China}%
	%\author{Jiajie Liang}
	%\affiliation{China Institute of Atomic Energy, P. O. Box 275(18), Beijing 102413, China}

\date{\today}

\begin{abstract}
%The inherent fluctuation relationship between the impact parameter and experimental observables is investigated within the framework of improved quantum molecular dynamics (ImQMD) model. Our calculations show that the intrinsic fluctuations are mainly generated in the microscopic stochasticity of initialization and nucleon-nucleon collisions in the nonequilibrium process of heavy ion collisions. Consequently, a given value of observable will correspond to a distribution of the input variables, such as impact parameter. 
To reconstruct the impact parameter distributions from the selected events sample or centrality, which is defined by two-observables, at intermediate energy heavy ion collisions, we extend the approach proposed by Das \textit{et al.} [Phys. Rev. C 97, 014905 (2018)], Rogly \textit{et al.} [Phys. Rev. C 98, 024902 (2018)], and Frankland \textit{et al.} [Phys. Rev. C 104, 034609 (2021)]. Based on deep investigations of the fluctuation mechanism, we found that the intrinsic fluctuations are mainly generated in the microscopic stochasticity of initialization and nucleon-nucleon collisions in the nonequilibrium process of heavy ion collisions, and this leads the observables to fluctuate with respect to impact parameter in a Gaussian form. In this work, the multiplicity of the charged particles and the total transverse momentum of the light charged particles are used simultaneously to model-independently reconstruct the impact parameter distributions for selected events or centrality based on the Bayesian method. For sorting the centrality with two observables, we propose to use the $K$-means clustering method (an unsupervised machine learning algorithm), which can automatically sort events when the class number is given. Furthermore, the reconstructed impact parameter distributions from data of the two observables can be used to learn the correlation between multiplicity and transverse momentum at different centralities, which may be useful for understanding the fragmentation mechanism.

%by using the multi-observable to reconstructing the impact parameter distributions model independently, the correlations among the observable pair can be learned. % Correspondingly, the impact parameter distribution can be obtained based on the Bayes' theorem
\end{abstract}

\maketitle

\section{Introduction}
Intermediate energy heavy ion collisions (HICs) provide a unique way to learn the equation of state (EoS) of bulk nuclear matter in the laboratory. In more detail, the strategy for learning EoS in the laboratory is to compare the data of selected collisions with the predictions of transport models. 
%However, the EoS cannot be obtained directly by measurement but has to be inferred by comparing the data of selected collisions with the predictions of transport models. 
To get a reliable constraint on the EoS, two aspects should be investigated or considered. One is to understand the uncertainties from transport models, which stimulates the transport model evaluation project (TMEP)\cite{Kolomeitsev05,Xujun16,HermannPPNP22,Maria21,YXZhang18,Akira19}, and some important progresses have been made in the treatment of the nucleonic mean field~\cite{Maria21} and the collision~\cite{YXZhang18,Akira19}. The other is to simulate the HICs with the same conditions as in experiments, for example, the same impact parameter distributions. This also stimulates studies on how to sort or estimate the impact parameter distributions for reducing the uncertainties due to the mismatch of experimental centrality in transport model simulations\cite{Bass1994,Christophe1995,Haddad1997,FPLi20,CYTSang21,Kuttan20}.

The impact parameter $b$ is not directly measurable and is usually estimated from a single observable or multiple observables with different methods. Generally, the methods for estimating impact parameters or reconstructing impact parameter distributions can be divided into three types~\cite{Lili2022}. The first one is the sharp cutoff approximation, which was proposed by Cavata \textit{et al.,}~\cite{Cavata1990} and has been widely used~\cite{Peter90,Phair1992,LUKASIK97,Andronic2006,TXLiu12}. The second one is the machine learning method, such as the artificial neural network(ANN) \cite{Bass1994,Christophe1995,Haddad1997}, convolutional neural network (CNN), light gradient boosting machine(LightGBM)~\cite{FPLi20,CYTSang21}, and PointNet models\cite{Kuttan20}. The third one is a model-independent method for reconstructing experimental impact parameter distributions, which was proposed by Das~\textit{et al.}\cite{Das18} and further developed in Refs.\cite{Rogly18, Frankland21,Yousefnia22}. In this paper, we refer to it as the Bayesian method.

The first and second methods assume that the observables have a one-to-one correspondence with $b$, and this idea has inspired a series of efforts to search for a way to accurately determine impact parameters. However, this assumption fails for intermediate energy HICs, %these works neglect the properties of fluctuation of observables to $b$ in  intermediate energy HICs. 
because the strong fluctuations of observables with respect to $b$ have been observed in experiments and transport model simulations\cite{Chomaz04,Gossiaux97,Phair1992,Ogilvie1989,Zhang04,Lili2018}. % It has been verified in the transport model simulations\cite{Phair1992,Ogilvie1989,Zhang04,Lili2018}. %For example, in the QMD simulations, a tiny difference among the initial states %(the relative difference of sampled nucleon's position and momentum are less than $10^{-7}$)
%leads to different values of final observables\cite{Zhang04}, and the chaocity is quantified by the Lyapunov exponent. 
Consequently, different values of the observables can coexist in simulations even for the same impact parameter. Conversely, the same value of the observable could correspond to the different impact parameters. But, these situations also raise a question of whether one can use as many observables as possible to determine $b$ uniquely. Otherwise, one should reconstruct impact parameter distributions from the HIC observables.% In the later situations, the corresponding simulations should be run over the same impact parameter distributions, or at least use a representative value such as the mean of \textcolor{red}{these} distributions, $\langle b\rangle$. %This behavior is also known as the impact parameter smearing effect\cite{Phair1992,Ogilvie1989,Lili2018}. 

The third method considers the fluctuation mechanism of the observables for $b$ and reconstructs the impact parameter distribution from a selected sample of events. This method is based on Bayes's theorem,
\begin{equation}
\label{Bayes1}
P(b|X)=P(b)P(X|b)/P(X).
\end{equation}
%The $c_b$ is named as $b$-centrality, and is defined as $c_b=\int_0^b P(b') db'$, with the probability distribution of $b$, $P(b)$.
$P(X)$ is the probability density of the observable $X$ which can be measured in the experiment, $P(X|b)$ is the probability density distribution of $X$ at given impact parameter $b$. The form of $P(X|b)$, also named as the fluctuation kernel~\cite{Frankland21}, is assumed to be a Gaussian \cite{Das18} or gamma distribution~\cite{Rogly18} for taking into account the fluctuation. Usually, the observable $X$ was chosen as the multiplicity of charged particles \cite{Das18,Rogly18,Frankland21}. The centroid and width of the Gaussian distribution, or the shape and scale of the gamma distribution are assumed in advance and they depend on $b$. The values of these parameters were determined by reproducing the experimental data of $P(X)$ with the formula $P(X)=\int P(X|b)P(b) db$. To avoid the uncertainties in the overall impact parameter distributions of $P(b)$, Das \textit{et al.} introduced $b$ centrality \cite{Das18}, i.e., $c_b=\int_0^b P(b') db'$, to replace the variable $b$. The replacement leads to $P(c_b)=1$, and $P(X)=\int P(X|c_b) dc_b$. By fitting the data of $P(X)$, one can find the solution of $P(X|c_b)$ and then $P(c_b|X)$ can be obtained based on Bayes's theorem. Then the expected impact parameter distributions of selected events can be retrieved from $P(b|X)=P(b)P(c_b|X)$\cite{Frankland21}. In those works, they mainly focused on how to obtain the impact parameter distribution model-independently with different forms of fluctuation kernel, but discussed less the origin of the fluctuation kernel in physics. Furthermore, one may expect to use the Bayesian method to reconstruct the impact parameter distribution from multiple observables, which may reveal the correlation between different observables as a function of centrality. A related issue has been discussed in Ref.\cite{Yousefnia22} for high energy HICs, but there is no work using this method in low-intermediate energy HICs. %, which will be useful for learning the reaction mechanism and constraining the equation of state through HICs. 

%his method has been used from ultra-relativistic collisions to low-intermediate energy collisions.

In this work, we investigate whether one can use as many observables as possible to uniquely determine $b$ by exploring the fluctuation mechanism within the framework of the improved quantum molecular dynamics (ImQMD) model \cite{Zhang2014,Zhang20FOP}. Then, we adopt the Bayesian method to reconstruct the impact parameter distributions from two observables, i.e., the multiplicity of charged particles $M$ and total transverse momentum of light particles, $p_t^{tot}$, for selected event samples or centrality. In addition, the uncertainties and bias of the reconstructed covariance matrix elements, which represent the fluctuation of the multiplicity and total transverse momentum of light particles, are discussed. For the selection of event samples in the multidimensional observables space, we propose to use an unsupervised machine learning algorithm, $K$-means, to automatically handle it. %By using the Bayesian method, the variance/covariance of \textcolor{red}{observables} for selected events or centrality are also obtained in this method, which may be useful for learning the fragmentation mechanism. %The two-dimensional fluctuation kernel $P(\mathbf{X}=\{M, p_t^{tot}\}|b)$ for model-independently reconstructing the impact parameter distributions are verified by the ImQMD model. The theoretical values of parameters in the fluctuation kernel can be used as a initial values for the experimentalist to do the model-independent reconstruction of impact parameter distributions from multiplicity and transverse momentum of light particles. 

%Very recently, Yousefnia \textit{et al.}

%Inspired by this work, we first check the origins of fluctuation in the ImQMD calculations by analyzing the influences of initialization, nucleonic mean field and nucleon-nucleon collisions on HIC observables, such as multiplicity of charged particles ($M$) and total transverse momentum of light particles ($p_t^{tot}$). 

%Before the study of the reconstruction of impact parameter distributions by using the Bayesian approach, 
\section{Fluctuation mechanism in the ImQMD model}
\label{fluctuation}
Now, let us investigate the origins of the fluctuation in HICs and why the impact parameter cannot be uniquely determined with as many observables of the HICs as possible. % based on the ImQMD calculations, as it is closely correlated with the impact parameter mixing. 

Theoretically, the fluctuation of final observables in HICs arises from the many-body correlation term in the transport equation. In the Boltzmann-Uehling-Uhlenbeck model, it can be realized by involving the fluctuation term \cite{FSZhang93,Maria98,Chomaz04,Maria13,WJXie13,Paolo13,Paolo15}. In the quantum molecular dynamics model, it can be realized by involving both the microscopic stochasticity of initialization and nucleon-nucleon collisions with the fixed width of the Gaussian wave packet. 

To quantitatively illustrate them in the framework of quantum molecular dynamics model, we perform the calculations of $^{112}$Sn+$^{112}$Sn at b=2 fm with the ImQMD model \cite{Zhang2014} under different strengths of initial fluctuations and nucleon-nucleon collisions. The different strengths of the initial fluctuation are realized by using two kinds of initialization, i.e., standard and perturbative initializations. The different strengths of nucleon-nucleon collisions are realized by choosing different bombarding energies, i.e., $E_{beam}$=50 and 120 MeV/u, and by switching on and off the nucleon-nucleon collisions (named the full and Vlasov modes in this paper). 

%The full mode means the calculations with both mean field potential and nucleon-nucleon collisions, while the Vlasov mode means the calculations with only the mean field. Comparing the results obtained with the two modes, one can learn the contributions from the nucleonic mean field and nucleon-nucleon collisions. Two initializations are adopted. 
The standard initialization means that the positions of nucleons are sampled within the radius of nuclei, and the momenta of nucleons are sampled within the Fermi momentum which depends on the local density. The initial nuclei are finally selected under the requirements of fitting the binding energy (for more details see Ref.\cite{Zhang20FOP}). In the ImQMD simulations, the HICs are simulated event by event and the initial nuclei of different events are different in microscopic states or in the 6-A dimensional phase space.
%the events are independent of each other since the initial nuclei are prepared completely random under the requirements of proper nuclear properties, such as the nuclear radius and binding energy. 
Quantitatively, we define a dimensionless distance between the first event and $k$th event in phase space as
\begin{equation}\label{distance}
	D_{1k}=\sqrt{\sum_{i=1}^{A}[\frac{(\mathbf{x}_1(i)-\mathbf{x}_k(i))^2}{R_0^2}+\frac{(\mathbf{p}_1(i)-\mathbf{p}_k(i))^2}{P_0^2}}],
\end{equation}
to describe the strength of the initial fluctuation between first and $k$th events. In Eq. (\ref{distance}), the radius of compound nuclei , i.e., $R_0=1.2(A_p+A_t)^{1/3}$ fm, and $P_0=0.263$ GeV/$c$ are used to normalize the coordinate and momentum to dimensionless variables. $A_p$ and $A_t$ are the numbers of nucleons of the projectile and the target nuclei, respectively. The summation in Eq. (\ref{distance}) runs over all nucleons in the system. For the standard initialization, the distribution of $D_{1k}$ has a Gaussian shape, and its averaged value $\langle D_{1k}\rangle$ and standard deviation (or the width of distribution) $\sigma_{D_{1k}}$ are about $18.0$ and $2.0$ with the normalization factors $R_0$ and $P_0$. 
The perturbative initialization means that the initialization between any two events has a very tiny difference in phase space. In this work, we set $D_{1k} <10^{-7} $ at the initial stage between the first and kth events, which is far less than the distance for standard initialization. %In the latter, the effects from the mean field or stochastic nucleon-nucleon collisions can be learned. 
By comparing the results obtained with two kinds of initialization, one can understand the fluctuation originated from initializations. 

\begin{figure}[htbp]
	\centering
	\includegraphics[angle=0,scale=0.35,angle=0]{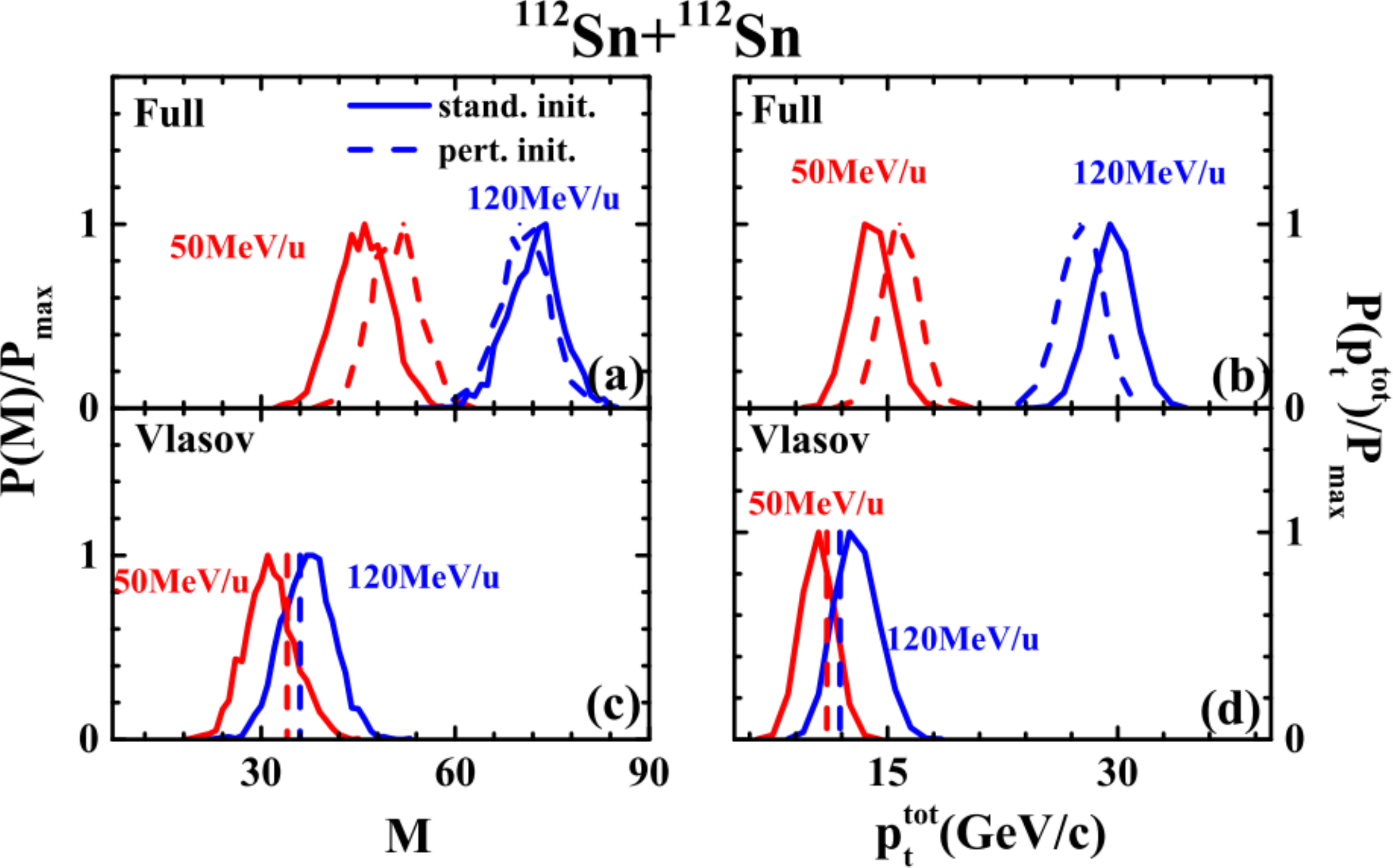}
	\setlength{\abovecaptionskip}{0pt}
	%\vspace{5em}
	\caption{ Height normalized distributions of $M$ and $p_{t}^{tot}$ obtained with the full mode [panels (a) and (b)] and Vlasov mode [panels (c) and (d)] under the conditions of standard initialization and perturbative initialization. The calculations are performed for $^{112}$Sn+$^{112}$Sn at b=2 fm.}
	\label{fig:fig1}
	\setlength{\belowcaptionskip}{0pt}
\end{figure}

We then perform the calculations by using two different initializations within the full and Vlasov modes, respectively. Figure \ref{fig:fig1} (a) and (c) show the height normalized distribution of the multiplicity of charged particles, i.e., $P(M)/P_{max}(M)$, in the full and Vlasov modes with two kinds of initialization. Two beam energies are simulated: one is 50 MeV/u (red lines) and the other is 120 MeV/u (blue lines). In the full mode, results obtained with both the standard initialization (solid lines) and perturbative initialization (dashed lines) show a Gaussian shape, but the widths of the distributions are different. As listed in Table ~\ref{tab:widths}, the widths of the distributions obtained with the standard initialization are larger than those with the perturbative initialization. In the case of the Vlasov mode, the multiplicity distributions for two kinds of initialization become completely different. As shown in panel (c), the multiplicity distributions in the case of the standard initialization still keep a Gaussian shape and have widths of about $\approx 20$, but the multiplicity distributions in the case of the perturbative initialization become $\delta$ distributions due to the absence of the large fluctuation caused by stochastic nucleon-nucleon collisions.

\begin{table}[htbp]
	\centering
	\caption {The widths of multiplicity distributions from standard and perturbative initializations in the cases of full and Vlasov modes. The numbers in brackets are the widths of distribution of total transverse momentum of light particles.}
	\begin{tabular}{c|*{3}{c}}
		\hline
		\hline
		{Mode}& $E_{beam}$(MeV/u) & Stand. Init. & Pert. Init. \\
		%		&\makebox[3em]{Standard}&\makebox[3em]{Perturbative}\\\hline%\hline
		\hline
		%		\hline
		Full & 50 & 4.13 (1.25) & 3.74 (1.17)  \\
		%		\hline
		Full & 120  & 4.13 (1.39) & 3.97 (1.35) \\
		Vlasov & 50  & 3.87 (1.17) & 0  (0) \\
		Vlasov & 120 & 3.99 (1.37) & 0  (0) \\
		
		\hline
		\hline
	\end{tabular}\label{tab:widths}
\end{table}

\begin{figure}[htbp]
	\centering
	\includegraphics[angle=0,scale=0.35,angle=0]{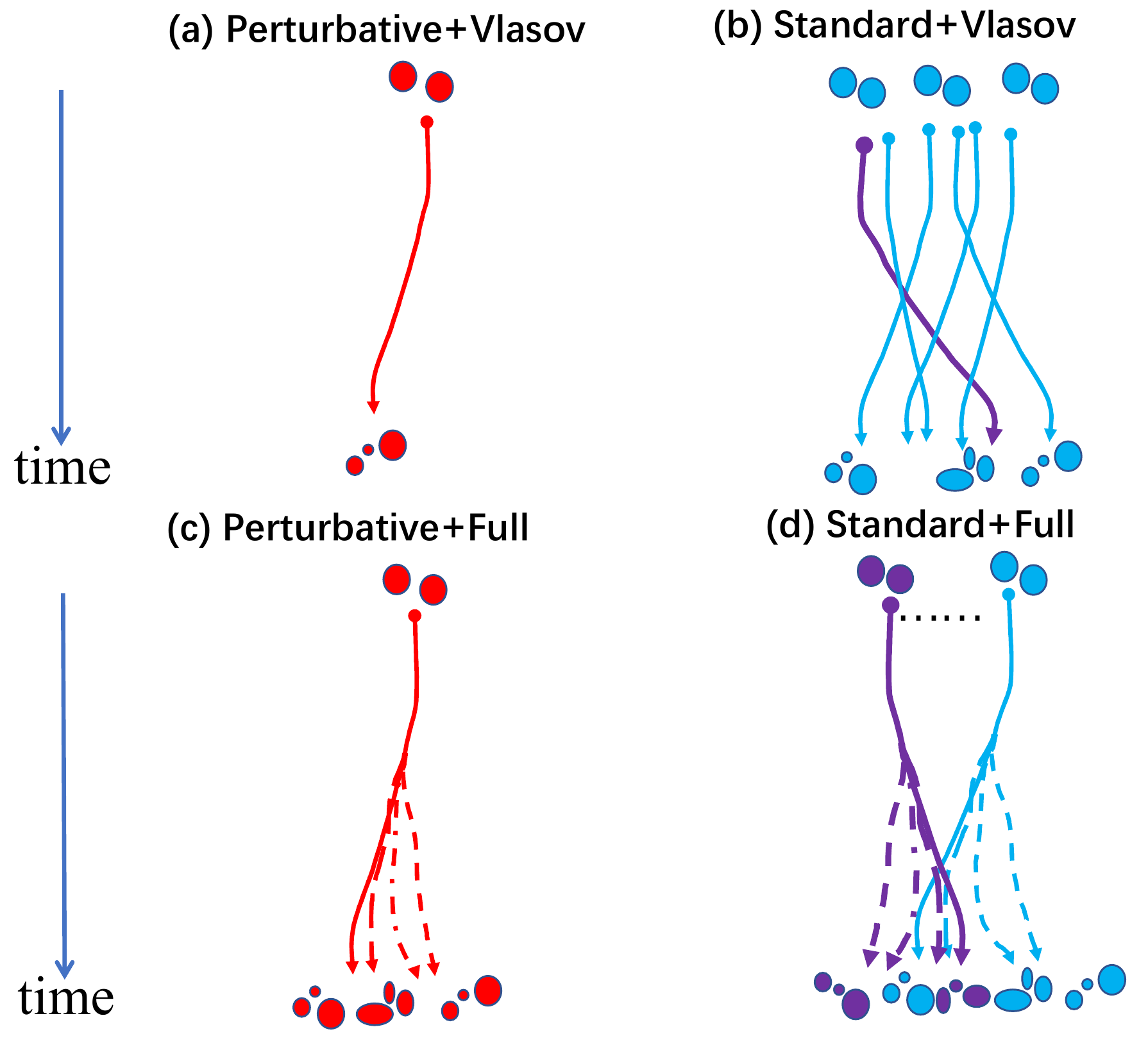}
	\setlength{\abovecaptionskip}{0pt}
	%\vspace{5em}
	\caption{Sketch of trajectories for different events in the cases of perturbative and standard initialization with Vlasov mode and full mode.}
	\label{fig:fig1-branch}
	\setlength{\belowcaptionskip}{0pt}
\end{figure}

The behaviors mentioned above can be understood from the philosophies of the QMD appoach, which are presented as a sketch in Fig.~\ref{fig:fig1-branch}. For convenience, let us start from the Vlasov mode. The lines in Figs.~\ref{fig:fig1-branch}(a) and ~\ref{fig:fig1-branch}(b) represent Vlasov trajectories of events for perturbative and standard initializations in phase space, respectively. In the Vlasov mode, the particles experience only the self-consistent effective mean field, so that the final observables are strongly correlated to the strength of the initial fluctuation. Consequently, the perturbative initialization leads to a $\delta$ distribution of final observables, as shown in Fig. \ref{fig:fig1}(c). However, a wide distribution of final observables from different events appears with the standard initialization, as shown in Fig. \ref{fig:fig1}(c), which is attributed to the large strength of fluctuation of the initialization. 

For the full mode, there is a wide distribution of final observables from different events even for the perturbative initialization. It comes from the various stochastic nucleon-nucleon collisions, and we depicted it as the dashed lines in Fig.~\ref{fig:fig1-branch}(c). In the standard initialization, both the initialization and stochastic nucleon-nucleon collision influence the distribution of final observables, which is illustrated in Fig.~\ref{fig:fig1-branch}(d). The widths of distributions of observables increase a little bit compared to the results of the perturbative initialization as shown in Table~\ref{tab:widths} because the trajectories of different events are independent in the QMD approach. 

The total transverse momentum distribution of light charged particles, i.e., $p_t^{tot}=\sum_{i} p_{t}(i)$, obtained by the summation of transverse momentum for light particles with $Z\le 2$, also shows a Gaussian-type distribution as shown in Figs. \ref{fig:fig1}(b) and \ref{fig:fig1}(d). The results from the full mode and Vlasov mode confirmed again the roles of the initialization and collisions in fluctuation.

Based on the above discussions, one can definitely draw a conclusion that accurate determination of the impact parameter is impossible even with as many HIC observations as possible. The reason is that the one-to-one correspondence between the final observables and the initial states is destroyed by the initial fluctuation and random nucleon-nucleon scattering.

\section{Probability density function $P(\mathbf{X}=\{M_0, p_{t0}^{tot}\}|b)$ from pseudo-event data}
To get the impact parameter distributions with the Bayesian method from two observables, $M$ and $p_{t}^{tot}$, one has to first determine the probability density function (PDF) of the observable vector $\mathbf{X}=\{M, p_t^{tot}\}$ at given impact parameter $b$, i.e., $f(\mathbf{X},b)=P(\mathbf{X}|b)$, named the fluctuation kernel as in Ref. \cite{Frankland21}. %In Refs.\cite{Das}, they proposed a gaussian distribution or a gamma distribution to describe the PDF of multiplicity of charged particles at given $b$ or $c_b$. Naturelly, one may expect to extend this method to use two observables, such as $M$ and $p_{t}^{tot}$, simultaneously.
In this work, two methods are used to extract the PDF from the pseudo data which is generated by the ImQMD model\cite{Zhang2014}. One is named direct calculation, which means calculating the distributions of the observables at given $b$, i.e., $P(\mathbf{X}|b)$, and thus is model dependent. Another is named the reconstructing method, which means to fitting the ``measured'' data of $P(\mathbf{X})$ to reconstruct $P(\mathbf{X}|b)$. The second method only needs the `measured' data of $P(\mathbf{X})$ without knowing the $b$ in advance, and thus is model independent. 

The calculations with the ImQMD model are performed for $^{112}$Sn+$^{112}$Sn at $E_{beam}$=120 and 50 MeV/u for generating the pseudodata. The pseudodata contain the information of the real impact parameter and can be used to check the validity of the second method. The number of events is 1,000,000, and the impact parameter $b$ is randomly distributed in the range from 0 to $b_{max}=1.2 (A_p^{1/3}+A_t^{1/3})$ %$b_{max}=1.2 (A_P^{1/3}+A_T^{1/3})$
fm according to the probability density $2b/b_{max}^2$.

\subsection{Direct calculation of $P(\mathbf{X}|b)$}
\begin{figure}[htbp]
	\centering
	\includegraphics[angle=0,scale=0.4]{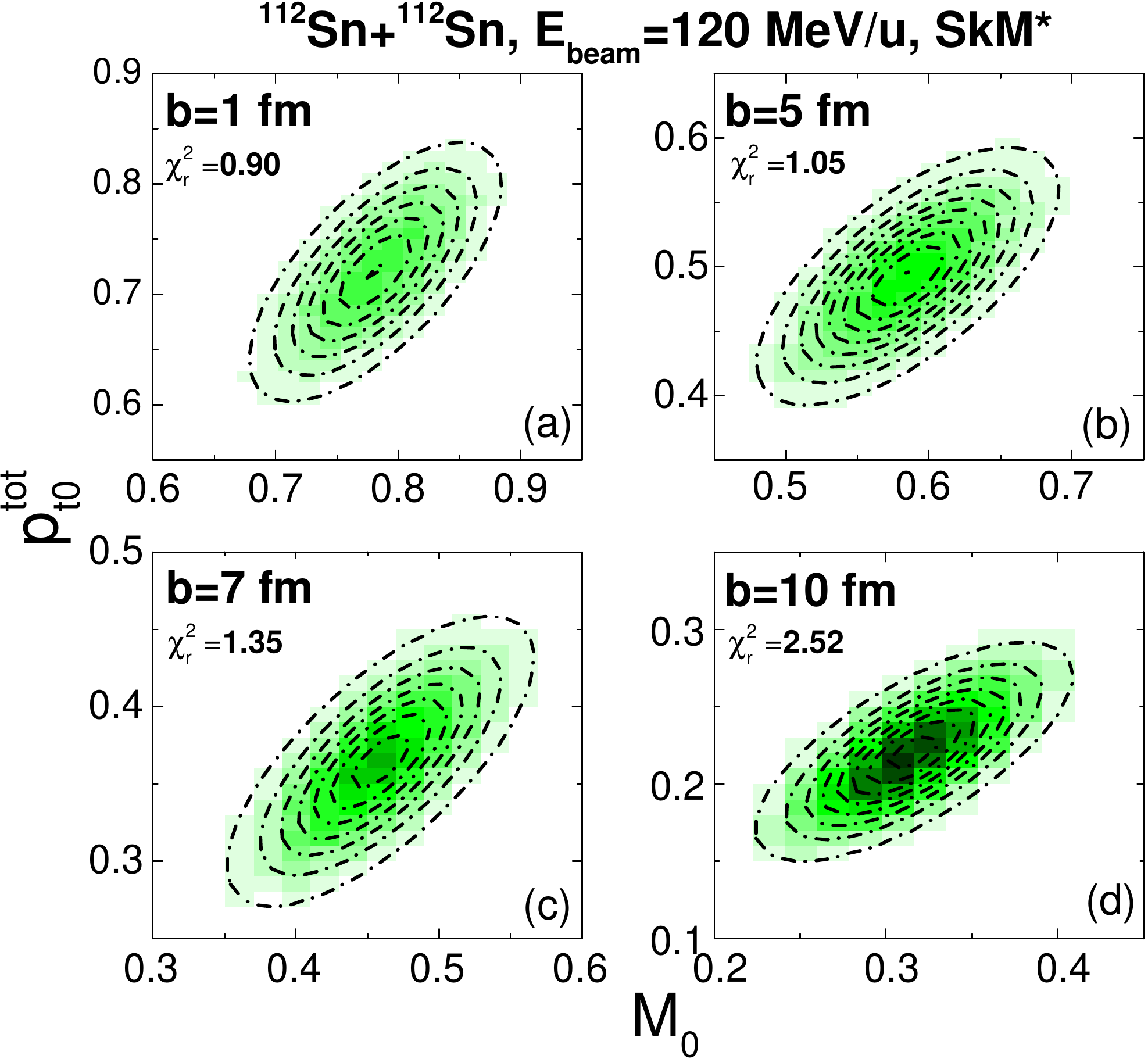}
	\setlength{\abovecaptionskip}{0pt}
	%\vspace{5em}
	\caption{Contour plots of probability density distribution $M_0$ vs $p_{t0}^{tot}$. Panel (a) is for 1 fm, (b) is for 5 fm,  (c) is for 7 fm, (d) is for 10 fm. The black dashed lines are the results obtained with the assumed PDF formula.}
	\label{fig:fig3-mpt-snsn}
	\setlength{\belowcaptionskip}{0pt}
\end{figure}

As an example, Fig.~\ref{fig:fig3-mpt-snsn} shows the contour plots of two observables distribution, i.e., P($M_0$,$p_{t0}^{tot}$) with $M_0=M/M_{max}$ and $p_{t0}^{tot}=p_{t}^{tot}/p_{t,max}^{tot}$, which is obtained with the ImQMD model for $^{112}$Sn+$^{112}$Sn at 120 MeV/u. The number of events at each $b$ is 60,000. $M_{max}$ and $p_{t,max}^{tot}$ are the maximum multiplicity of charged particles and the maximum total transverse momentum of light charged particles in the calculations, respectively. The values of them in our calculations can be found in Table~\ref{tab:max}. The panels (a), (b), (c) and (d) are the results obtained at $b$=1, 5, 7, and 10 fm, respectively. The two-dimensional PDFs of $\mathbf{X}=\{M_0, p_{t0}^{tot}\}$ distribute as a Gaussian shape. %, which is also viewed as a consequence of central limit theorem\cite{Rogly18}. %For very central and very peripheral collisions, the two-dimensional PDFs deviate from gaussian shape.

Except for the ImQMD simulations, the selection of Gaussian form of the PDFs is also a result of probability theory. As we know, the particles are detected with probability $p$ or not with probability $1-p$ for one event in experiments. It leads to a binomial distribution of observables. If the number of events is large enough, the binomial distribution tends to a Gaussian distribution according to the central limit theorem. 
%
%Another choice is gamma distribution. Gamma distribution arises naturally in which the waiting time between Poisson distributed events are relevant to each other. This is also similar with the particle detected by detector during the running time. 

%Based on this idea, I think the filter will not change the distribution, but will change the parameter in this distribution.

\begin{table}[htbp]
	\caption {Maximum multiplicity of charged particles and total transverse momentum of light charged particles used in %system $^{40}$Ca+$^{40}$Ca, $^{112}$Sn+$^{112}$Sn, and $^{197}$Au+$^{197}$Au for different beam energies.}
        system $^{112}$Sn+$^{112}$Sn for different beam energies.}
	\begin{tabular}{ccccccccccccccccccc}
		
		\hline
		System  &$E_{beam}$ (MeV/u)  &$M_{max}$   &$p_{t,max}^{tot}$ (GeV/$c$)\\
		\hline
%		&50  &36 &7\\
%		$^{40}$Ca+$^{40}$Ca     &120 &38 &11\\
%		&200 &40 &17\\
%		\\
		$^{112}$Sn+$^{112}$Sn  &50  & 61 &16\\
		 &120 & 86 &30\\
%		&200 &94 &46\\
%		&50  &82 &24\\
%		$^{197}$Au+$^{197}$Au   &120 &119 &47\\
%		&200 &140 &78\\
		\hline
		
	\end{tabular}
	\label{tab:max}
\end{table}

Based on previous discussions, one can assume a two-dimensional Gaussian form of the PDF of $M_0$ and $p_{t0}^{tot}$ as  %. In practical calculations, we adopt \textcolor{red}{the} two-dimensional \textcolor{red}{Gaussian}-shape PDF,
\begin{equation}
\label{GS-PDF}
P(\mathbf{X}|b)=\frac{\exp\{-\frac{1}{2}(\mathbf{X}-\overline{\mathbf{X}}(b))^T\Sigma^{-1}(b)(\mathbf{X}-\overline{\mathbf{X}}(b))\}}{2\pi\sqrt{|\Sigma(b)|}}.
\end{equation}
%\begin{eqnarray}
%	\label{2d-pdf}
%	&&P(\mathbf{X}=\{M_0,p_{t0}^{tot}\}|b)   \\\nonumber
%	&=&A\exp[-\frac{1}{2}(\frac{(M_0-\overline{M}_0)cos\theta+(p_{t0}^{tot}-\overline{p}_{t0}^{tot})sin\theta}{w_1})^2 \\\nonumber
%	&&-\frac{1}{2}(\frac{(M_0-\overline{M}_0)sin\theta-(p_{t0}^{tot}-\overline{p}_{t0}^{tot})cos\theta}{w_2})^2 ],
%\end{eqnarray}
%is relatively easy to fit the data. 
$\overline{\mathbf{X}}=\{\overline{M}_0, \overline{p}_{t0}^{tot}\}$ is the mean value of $\mathbf{X}$, and $\Sigma$ is the symmetric covariance matrix. $\Sigma^{-1}$ denotes the inverse matrix and $|\Sigma|$ is the determinant. 
%The $\overline{M}_0$ and $\overline{p}_{t0}^{tot}$ correspond to the mean values of $M_0$ and $p_{t}^{tot}$ in the case of complete guassian shape of PDF. $A$, $\theta$, $w_1$, and $w_2$ are the 4 parameters of PDF. $A$ is the height of gaussian functions, and its values depend on $\theta$, $w_1$, and $w_2$. 
%In this work, we normalize the distribution which is obtained in the whole impact parameter region. The width parameters $w_1$, $w_2$ and parameter $\theta$ are related to the variance of $M_0$ and $p_{t0}^{tot}$ and its covariance, i.e., $\langle(M_0-\overline{M}_0 )^2 \rangle$, $\langle(p_{t0}^{tot}-\overline p_{t0}^{tot})^2 \rangle$ and $\langle(M_0-\overline{M}_0 )(p_{t0}^{tot}-\overline p_{t0}^{tot})\rangle$. % which is 2 times of the standard deviation of $M_0$, $w_1$=$2\sigma_{M_0}$;
By fitting the ImQMD results at different $b$ with Eq.(\ref{GS-PDF}), one can get the form of the PDF (dashed lines in Fig.~\ref{fig:fig3-mpt-snsn}) and the $b$ dependence of  $\overline{\mathbf{X}}$, $\Sigma$. This is named the direct fitting calculation.
%$A$, $\theta$, $w_1$, and $w_2$. % the $\Sigma$ matrix. 
%In this work, the quality of the fits is described by the reduced chi-square, i.e., $\chi^2_r$. 
The reduced chi-square $\chi^2_r$ values of the fitting at different impact parameters for $E_{beam}$=120 and 50 MeV/u are listed in Table \ref{tab:chi2}, and the values of $\chi_r^2\le 3$ over the whole range of impact parameters. The form of Gaussisan shape becomes relatively worse at very peripheral collisions due to the limitation of the range of impact parameter. 

%$\overline{M}_0$, $\overline{p}_{t0}^{tot}$, 

\begin{table}[htbp]
	\caption {The $\chi^2_r$ of each fitting at different impact parameters for $^{112}$Sn+$^{112}$Sn and $E_{beam}$=120 and 50 MeV/u.}
	\begin{tabular}{cccccccccccc}
		
		\hline
		\hline
		& b (fm) & 0.0 & 1.0 & 2.0 & 3.0  & 4.0 & 5.0    \\
	120 MeV/u &	$\chi^2_r$ & 0.96 & 0.90 & 1.01& 1.01& 1.07 & 1.05\\
        50 MeV/u &	$\chi^2_r$ & 0.96 & 1.06 & 1.01& 1.12& 1.13 & 1.04\\
		\hline
		 & b (fm) & 6.0 & 7.0 & 8.0 & 9.0 & 10.0 & 11.0 \\	
		120 MeV/u & $\chi^2_r$ & 1.27 &1.35 & 1.50 & 1.85& 2.52 & 2.89\\
        50 MeV/u & $\chi^2_r$ & 1.07 &1.07 & 1.16 & 1.20& 1.34 & 1.34\\
		\hline
		\hline
		
	\end{tabular}
	\label{tab:chi2}
\end{table}

%The parameters of $\theta$, $w_1$, and $w_2$ in Eq.(\ref{2d-pdf}) are related to the variance and covariance of $M_0$ and $p_{t}^{tot}$ distributions. Their relations can be obtained by describing the Eq.(\ref{2d-pdf}) with the general form of two-dimensional gaussian PDF,
%\begin{equation}
%\label{GS-PDF}
%P(\mathbf{X}=\{M_0,p_{t0}^{tot}\}|b)=\frac{\exp\{-\frac{1}{2}(\mathbf{X}-\bar{\mathbf{X}})^T\Sigma^{-1}(b)(\mathbf{X}-\bar{\mathbf{X}})\}}{2\pi\sqrt{|\Sigma|}}.
%\end{equation}
%In Eq.(\ref{GS-PDF}), $\bar{\mathbf{X}}$ is the mean value of $\mathbf{X}$, and $\Sigma$ is the symmetric covariance matrix. $\Sigma^{-1}$ denotes the inverse matrix and $|\Sigma|$ is the determinant. We finally have the following relation between the $\Sigma$ and four parameters of $A$, $\theta$, $w_1$, and $w_2$, 
%\begin{eqnarray}
%	\Sigma^{-1}_{11}(b)&=&\frac{\cos^2\theta}{w_1^2}+\frac{\sin^2\theta}{w_2^2},\label{s11-r}\\
%	\Sigma^{-1}_{22}(b)&=&\frac{\sin^2\theta}{w_1^2}+\frac{\cos^2\theta}{w_2^2},\label{s22-r}\\
%	\Sigma^{-1}_{12}(b)&=&\Sigma^{-1}_{21}(b)=\sin\theta\cos\theta(\frac{1}{w_1^2}-\frac{1}{w_2^2}).\label{s12-r}\\
%		A&=&\frac{1}{2\pi\sqrt{|\Sigma|}}\label{Asig}
%\end{eqnarray}

\begin{figure}[htbp]
	\centering
	\includegraphics[angle=0,scale=0.45,angle=0]{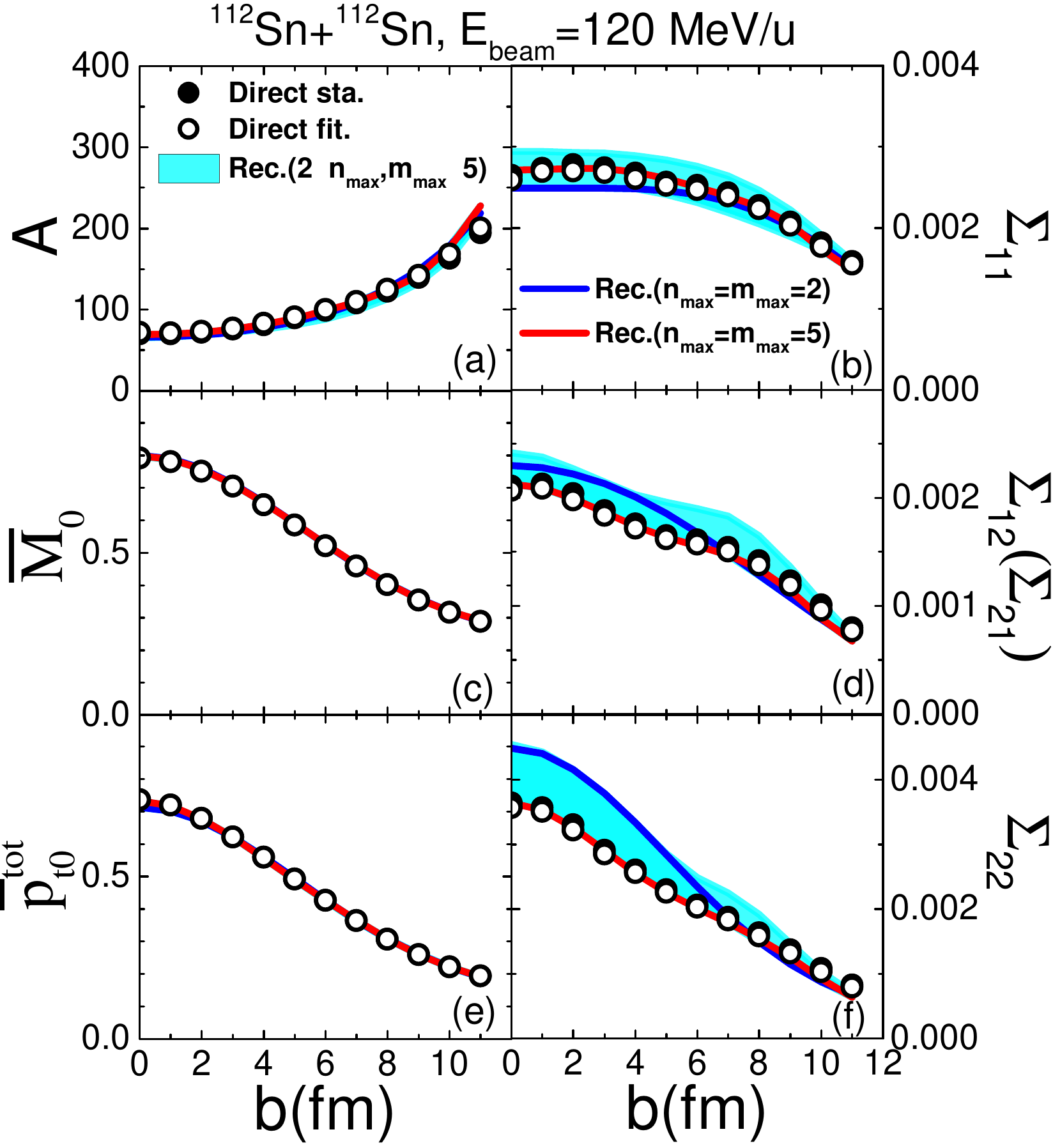}
	\setlength{\abovecaptionskip}{0pt}
	%\vspace{5em}
	\caption{The parameters $A$, $\overline{M}_0$, $\overline{p}_{t0}^{tot}$, $\Sigma_{11}$, $\Sigma_{12}(\Sigma_{21})$ and $\Sigma_{22}$ as functions of $b$ for $^{112}$Sn+$^{112}$Sn at 120 MeV/u. Solid circles are for the direct statistical calculation, the open circles are for the direct fitting calculation, and the cyan shaded regions and the blue and red lines are for the reconstructing method.
% under initialization parameters ($\overline{M}_0$, $\overline{p}_{t0}^{tot}$, $\Sigma_{11}$, $\Sigma_{12}(\Sigma_{21})$ and $\Sigma_{22}$) 
% for $^{112}$Sn+$^{112}$Sn at $E_{beam}$=120 MeV/u.
}
	\label{fig:fig4-fb}
	\setlength{\belowcaptionskip}{0pt}
\end{figure}

\begin{figure}[htbp]
	\centering
	\includegraphics[angle=0,scale=0.45,angle=0]{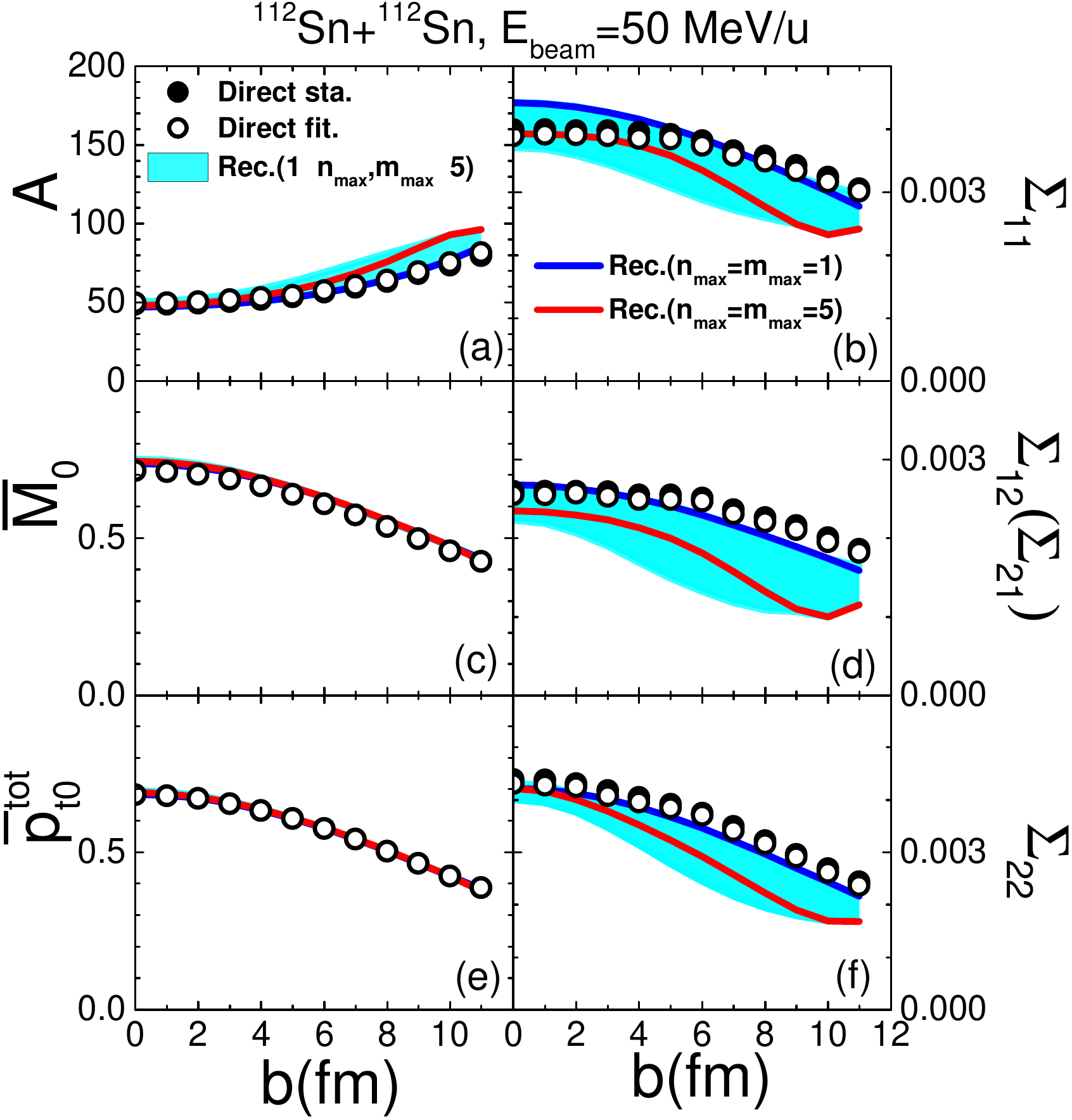}
	\setlength{\abovecaptionskip}{0pt}
	%\vspace{5em}
	\caption{Same as Fig.~\ref{fig:fig4-fb}, but for $E_{beam}$=50 MeV/u. }
	\label{fig:fig4-fb-E50}
	\setlength{\belowcaptionskip}{0pt}
\end{figure}

In Figs.~\ref{fig:fig4-fb}(a)-\ref{fig:fig4-fb}(f), we present $A(=1/(2\pi\sqrt{|\Sigma|}))$, $\overline{M}_0$, $\overline{p}_{t0}^{tot}$, $\Sigma_{11}$, $\Sigma_{12}(\Sigma_{21})$, and $\Sigma_{22}$ as functions of $b$ for $^{112}$Sn+$^{112}$Sn at $E_{beam}$=120 MeV/u. Figure~\ref{fig:fig4-fb-E50} shows the similar results at incident energy $E_{beam}$=50 MeV/u. The open circles are obtained from the direct fitting calculations.  % by fitting the Eq.(\ref{GS-PDF}) to ImQMD results at different $b$, and they are close to each other. 
%The points in panels (a)-(f) are obtained from $\overline{M}_0$ and $\overline{p}_{t0}^{tot}$, $A$, $\theta$, $w_1$, and $w_2$, by using Eq.(\ref{s11-r})-(\ref{Asig}). 
 $\overline{M}_0$ and $\overline{p}_{t0}^{tot}$ decrease with increasing impact parameter due to the decrease of the size of overlap region and the nucleon-nucleon collision frequency with increasing impact parameter. %It suggests that the centroid of \textcolor{red}{the} PDF or the mean values of observables may be used to estimate the impact parameter. 
 In addition, $\Sigma_{11}$, $\Sigma_{12}(\Sigma_{21})$, and $\Sigma_{22}$ decrease with increasing impact parameter, which reflects the decrease in the strength of the fluctuation due to the decrease in the nucleon-nucleon collision rate. 

The solid circles in panels (a)-(f) are obtained from the mean values, variance, and covariance of observables from their distributions at different $b$, i.e., by using the following equations:
\begin{eqnarray}
	\label{sigma-var}
        \overline{M}_0(b) &=& \langle M_0(b)\rangle, \\
        \overline p_{t0}^{tot}(b) &=& \langle p_{t0}^{tot}(b)\rangle,\\
	\Sigma_{11}(b)&=&\langle(M_0-\overline{M}_0 )^2 \rangle\label{sg11},\\
	\Sigma_{22}(b)&=&\langle(p_{t0}^{tot}-\overline p_{t0}^{tot})^2 \rangle\label{sg22},\\
	\Sigma_{12}(b)&=&\Sigma_{21}(b)=\langle(M_0-\overline{M}_0 )(p_{t0}^{tot}-\overline p_{t0}^{tot})\rangle\label{sg12}.
\end{eqnarray}
$\langle\cdot\rangle$ means an average over the events with the same impact parameter $b$. We named this method direct statistical calculations. The values obtained from Eqs.(\ref{sigma-var}-\ref{sg12}) can validate the applicability of the Gaussian PDF and direct fitting method. It is shown in Figs.~\ref{fig:fig4-fb} and ~\ref{fig:fig4-fb-E50} that the open circles are very close to the solid circles. The cyan shaded regions are obtained from the reconstructing method by directly fitting the data of $P(\mathbf{X})$, and we will discuss it in Sec.~\ref{subsec:Reconst}.

\subsection{Reconstructing $P(\mathbf{X}|b)$}
\label{subsec:Reconst}
%$P(X)=\int P(X|c_b)P(c_b) d c_b$
To reconstruct the impact parameter distribution model-independently, we adopt the formula
\begin{equation}
\label{PX-2d}
    P(\mathbf{X})=\int_0^1 P(\mathbf{X}|c_b)P(c_b)dc_b=\int_0^1 P(\mathbf{X}|c_b)dc_b,
\end{equation}
to fit the data of $P(\mathbf{X})$. %In Eq.(\ref{PX-2d}), $P(c_b)=1$ when we change the $b$ to \textcolor{red}{the} variable $c_b$ as same as in Ref.\cite{Das18,Yousefnia2022} %\cite{Yousefnia2022} for avoiding the uncertainty from impact parameter distributions. 
In our calculations, the form of $P(\mathbf{X}|c_b)$ is assumed to be
\begin{equation}
\label{GS-PDF-cb}
P(\mathbf{X}|c_b)=\frac{\exp\{-\frac{1}{2}(\mathbf{X}-\overline{\mathbf{X}}(c_b))^T\Sigma^{-1}(c_b)(\mathbf{X}-\overline{\mathbf{X}}(c_b))\}}{2\pi\sqrt{|\Sigma(c_b)|}}.
\end{equation}
The mean values $\overline{\mathbf{X}}$ and the elements of the covariance matrix $\Sigma_{ij}$ are smooth positive functions of $c_b$, and are expressed as the exponential of a polynomial as in Ref. \cite{Yousefnia22},
\begin{eqnarray}
	\label{mean-var}
	\overline{X}_i(c_b)&=&\overline{X}_i(0)\exp\left (-\sum_{n=1}^{n_{\mathrm{max}}} a_{i,n} c_b^n   \right ) \\
	\Sigma_{ij}(c_b)&=&\Sigma_{ij}(0)\exp\left (-\sum_{m=1}^{m_{\mathrm{max}}} A_{ij,m} c_b^m\right )
\end{eqnarray}
where $\overline{X}_i(0)$, $a_{i,n}$, $\Sigma_{ij}(0)$, $A_{ij,m}$ are free parameters, and $n_{\mathrm{max}}$ and $m_{\mathrm{max}}$ are the degrees of the polynomials used to parametrize the mean and the covariance. These parameters are adjusted to obtain the best fit of $P(\mathbf{X})$ by using the code MINUIT. %The parameters are fitted in such a way that the distribution (\ref{PX-2d}) matches data, and the degree of the polynomial is adjusted so as to obtain a satisfactory fit. 
%We carry out a standard χ2 fit, keeping all nonempty boxes in the histogram of (ET ,Nch ).
%Then, we fit the $P(\mathbf{X})$ data by Eq.~\ref{PX-2d} to reconstruct $P(\mathbf{X}|c_b)$ (or, $P(\mathbf{X}|b)$). The parameter $\overline{X}_i(0)$, $a_{i,n}$, $\Sigma_{ij}(0)$, $A_{ij,m}$ and the degree of the polynomial is adjusted so as to obtained a satisfactory fit. 

\begin{figure}[htbp]
	\centering
	\includegraphics[angle=0,scale=0.4,angle=0]{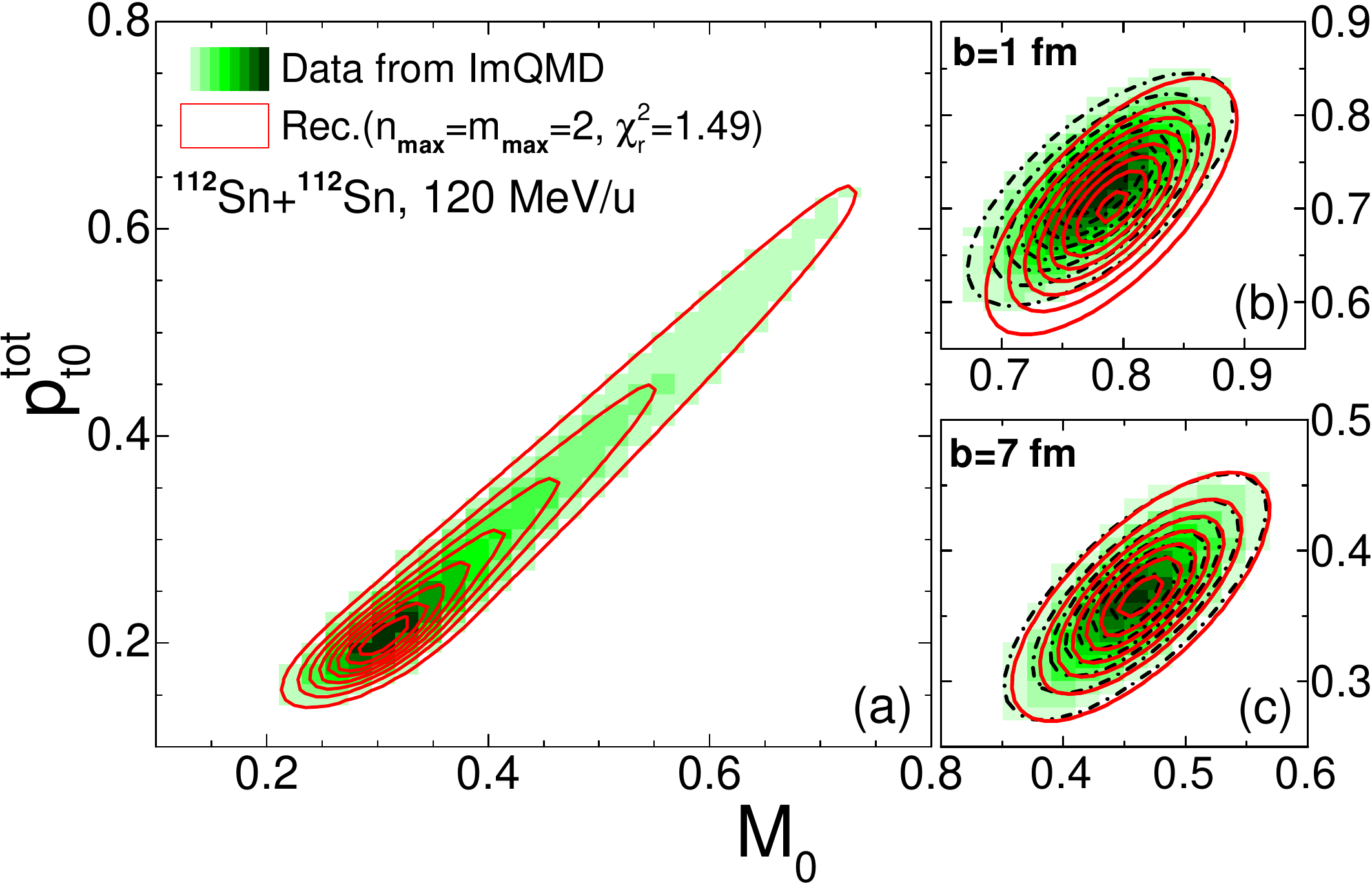}
	\setlength{\abovecaptionskip}{0pt}
	%\vspace{5em}
	\caption{Contour plots of probability density distribution $M_0$ vs $p_{t0}^{tot}$ for $^{112}$Sn+$^{112}$Sn at $E_{beam}$=120 MeV/u. The red solid lines are the results obtained from the reconstructing method: the black dashed lines are from the direct fitting calculation.  }
	\label{fig:fig5-fb}
	\setlength{\belowcaptionskip}{0pt}
\end{figure}

\begin{figure}[htbp]
	\centering
	\includegraphics[angle=0,scale=0.4,angle=0]{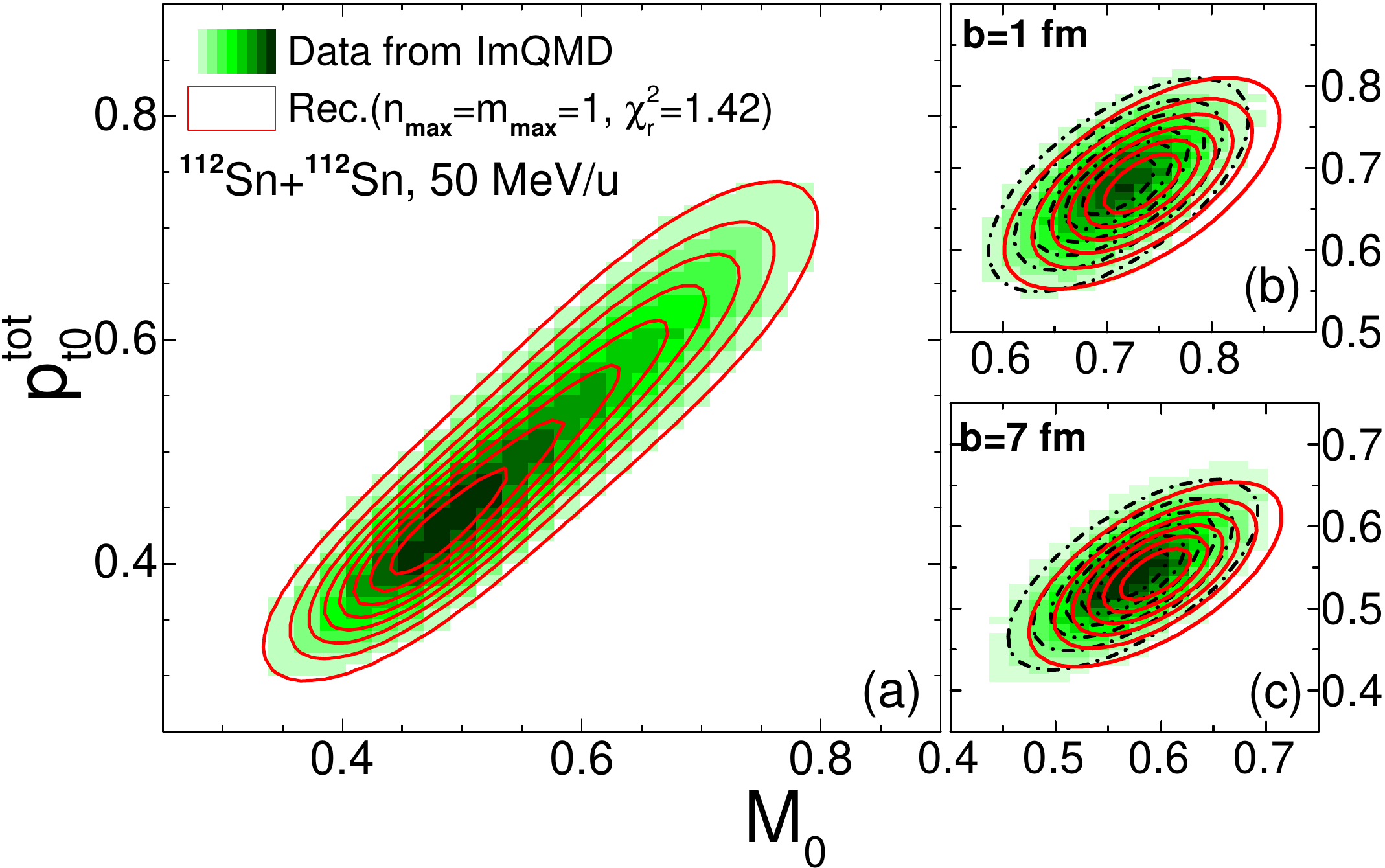}
	\setlength{\abovecaptionskip}{0pt}
	%\vspace{5em}
	\caption{Same as Fig.~\ref{fig:fig5-fb}, but for $E_{beam}$=50 MeV/u. }
	\label{fig:fig5-fb-E50}
	\setlength{\belowcaptionskip}{0pt}
\end{figure}

To directly view the validity of the reconstructing method, we first present the contour plot of $P(\mathbf{X})$ obtained with the ImQMD model (color map) and reconstructing method (red solid lines), which corresponds to the minimum fitting parameters when $\chi_r^2<$2 in Figs. \ref{fig:fig5-fb}(a) and \ref{fig:fig5-fb-E50}(a). The panels (b)-(c) in Figs.\ref{fig:fig5-fb} and \ref{fig:fig5-fb-E50} are contour plots of $P(\mathbf{X}|b)$, obtained by the direct fitting calculations (black dashed lines), and by reconstructing method at $b=1$ fm and $b=7$ fm, respectively. The reconstructing method can well reproduce both the data and the results from the direct fitting calculation when $b\geq7$ fm at 120 MeV/u. For central collisions, the reconstructed $P(\mathbf{X}|b)$ slightly deviates from the data along the $p_{t0}^{tot}$ direction. %The difference come from the overestimation of $\Sigma_{22}$ as shown in Fig.\ref{fig:fig4-fb} (f). 
At incident energy of $E_{beam}$=50 MeV/u, as shown in Figs.~\ref{fig:fig5-fb-E50}(b) and ~\ref{fig:fig5-fb-E50}(c), the reconstructing method can reproduce the shape of $P(\mathbf{X}|b)$ but the mean $M_0$ value of the Gaussian form deviates from the real value less than 5\%.
% differs from that from direct calculation and the difference becomes smaller, or even disappears ($b > 5$ fm), with the impact parameter increasing. 

The key point in the reconstruction is to find a reasonable number of the degrees of the polynomials, i.e., $n_{max}$ and $m_{max}$. When $n_{max}$ and $m_{max}$ are too small, the Bayesian method may not reproduce $P(\mathbf{X})$. Conversely, when the $n_{max}$ and $m_{max}$ are too large, one may confront an over-fitting issue. In experiments, it is hard to justify how many fitting parameters are good enough by only seeking the minimum of $\chi^2_r$ among the different parameter sets, since the real $b$ dependence of $\overline{X}_i$ and $\Sigma_{ij}$ or the real $b$ distribution are not known in advance. Consequently, we need to learn the uncertainties of the reconstructed results by using different combinations of $n_{max}$ and $m_{max}$, and the deviation (or bias) relative to the true values. % by comparing the results obtained by the transport models.}

For $^{112}$Sn+$^{112}$Sn at $E_{beam}$=120 MeV/u, $n_{max}=m_{max}=1$ can not reproduce $P(\mathbf{X})$ and the corresponding $\chi_r^2$ is about 23. When $2\le n_{max}\le 5$ and $2\le m_{max}\le 5$, the obtained $\chi^2_r$ values are in the range 1.26-1.64. $n_{max}>5$ and $m_{max}>5$ were not used due to the number of fitting parameters exceeding the limit of the MINUIT. In Fig.\ref{fig:fig4-fb}, $\overline{X}_i(b)$ and $\Sigma_{ij}(b)$ obtained with $2\le n_{max}\le 5$ and $2\le m_{max}\le 5$ are presented as cyan-shaded regions, which reflect the uncertainties caused by different choices of $n_{max}$ and $m_{max}$. The uncertainties of $\overline{X}_i$ are less than 1.4\%, and the uncertainties of $\Sigma_{ij}$ are less than 14\%.

\begin{figure}[htbp]
	\centering
	\includegraphics[angle=0,scale=0.37,angle=0]{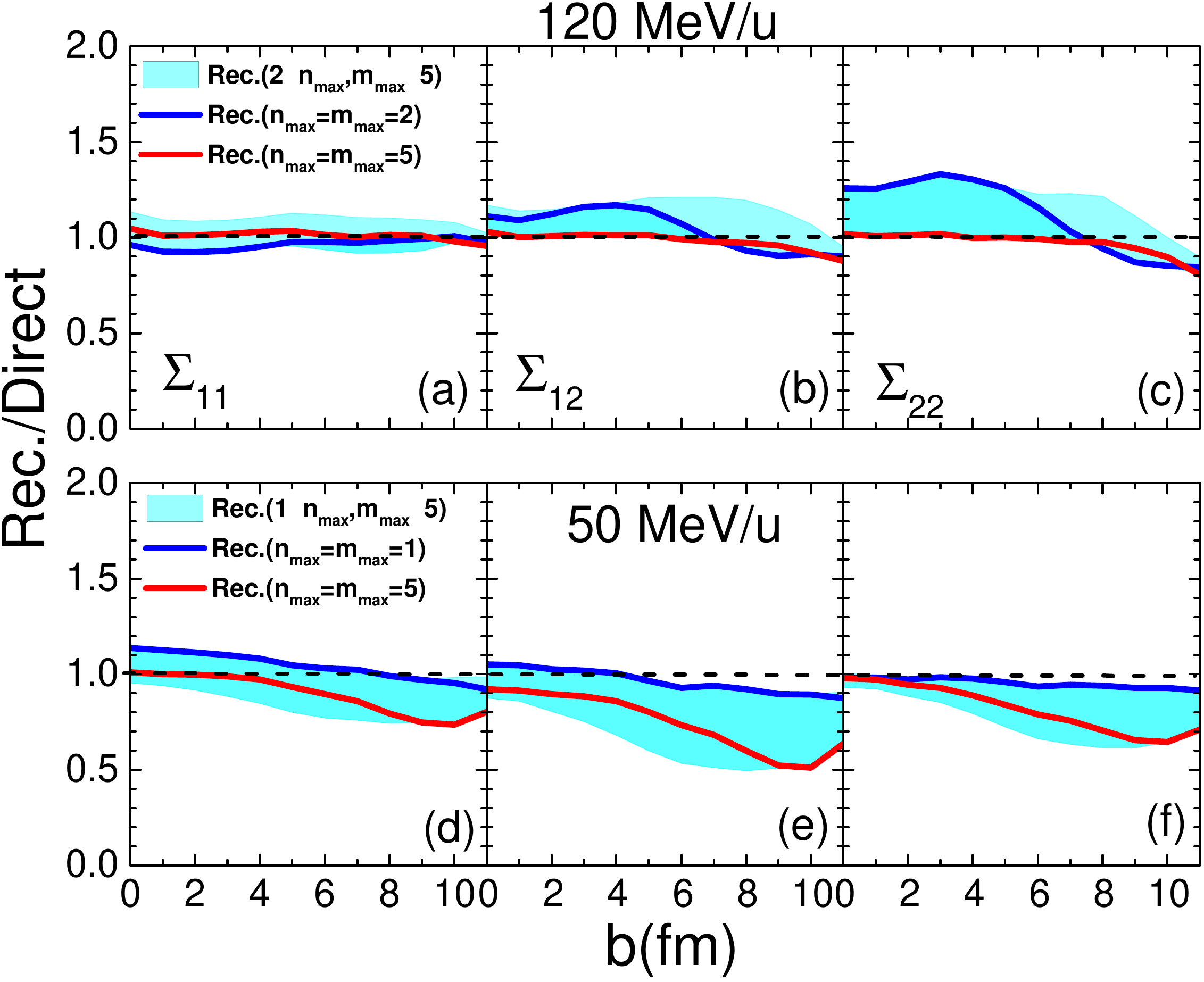}
	\setlength{\abovecaptionskip}{0pt}
	%\vspace{5em}
	\caption{The ratio of the reconstructing method and direct fitting calculation about $\Sigma_{11}$, $\Sigma_{12}$, and $\Sigma_{22}$ as functions of $b$ for $^{112}$Sn+$^{112}$Sn at 120 (upper panels) and 50 (lower panels) MeV/u.
% under initialization parameters ($\overline{M}_0$, $\overline{p}_{t0}^{tot}$, $\Sigma_{11}$, $\Sigma_{12}(\Sigma_{21})$ and $\Sigma_{22}$) 
% for $^{112}$Sn+$^{112}$Sn at $E_{beam}$=120 MeV/u.
}
	\label{fig:Fig8-Sigma-Rec_Dir}
	\setlength{\belowcaptionskip}{0pt}
\end{figure}

%The bias is presented by the deviation between the average values of the reconstructed fitting parameters, i.e., $\left<O\right>=\sum O_{rec.}(n_{max},m_{max})/N$ with $O=\overline{X}_i$ and $\Sigma_{ij}$, and true values. The summation in the expression of $\left<O\right>$ is over all combinations of $n_{max}$ and $m_{max}$ we used, and $N$ represents the number of combinations. Our calculations show that the deviations of averaged $\left<\overline{X}_i\right>$ from the direct fitting results are greater than 0.06\% but less than 2\%. 
In Figs. \ref{fig:Fig8-Sigma-Rec_Dir}(a)-\ref{fig:Fig8-Sigma-Rec_Dir}(c), the deviations of $\Sigma_{ij}$ at 120 MeV/u are presented by using the ratios between the reconstructed fitting parameters and direct fitting parameters. The blue lines are the results obtained with $n_{max}=m_{max}=2$, which is selected based on the maxim that fewer parameters are preferred than more if all of them can fit the data. In this case, the deviations of $\Sigma_{11}$ is less than 10\%, and the large deviations appear at $b<$5 fm for $\Sigma_{12}$ and $\Sigma_{22}$ and its values reach 33\%. %The $\chi_r^2$ of $n_{max}=m_{max}=2$ is 1.49. 
The red lines are the results obtained with $n_{max}=m_{max}=5$, and the reconstructing method ($n_{max}=m_{max}=5$) can reproduce the results of direct calculation. The slight deviations appears at very peripheral collisions.
%The reconstructing method can reproduce the results of direct calculation, except for $\Sigma_{22}$ which are slightly overestimated relative to the results of direct calculation at central collisions.  the selected combinations of $n_{max}$ and $m_{max}$,

For $^{112}$Sn+$^{112}$Sn at $E_{beam}$=50 MeV/u, the reconstructed $\Sigma_{ij}$ are presented in Fig.~\ref{fig:fig4-fb-E50}. The cyan-shaded region corresponds to the uncertainties obtained with $1\le n_{max}\le 5$ and $1\le m_{max}\le 5$, where the values of $\chi^2_r$ are in the range 1.03-1.42. The uncertainties of $\overline{X}_i$ are less than 1\%, and the uncertainties of $\Sigma_{ij}$ are less than 31\%. %The deviations of averaged $\left<\overline{X}_i\right>$ from the direct fitting results are greater than 0.01\% but less than 4.2\%. 
In Fig. \ref{fig:Fig8-Sigma-Rec_Dir}(d)-\ref{fig:Fig8-Sigma-Rec_Dir}(f), the deviations of $\Sigma_{ij}$ are presented. When $n_{max}\ge 2$ and $m_{max}\ge 2$, the reconstructed $\Sigma_{ij}$ are lower than that from those of the direct calculations. The reconstructed $\Sigma_{ij}$ with $n_{max}=m_{max}=1$ (blue lines) are close to that obtained with the direct calculations within 14\%, and the deviations appear at both central and peripheral collisions. The red lines are the results obtained with $n_{max}=m_{max}=5$. At $b<3$ fm, the deviations are less than 12\%. With the impact parameter increasing, the deviations increase and the largest deviation occur around $b$=9.5 fm.

%\textcolor{red}{Similar as in Ref.\cite{Yousefnia22}, the reconstructing method can reproduce the mean values of multiplicities and transverse momentum. But the reconstructed $\Sigma_{ij}$ may deviate from the direct calculation results, and the deviations depend on the choice of the selection of fitting parameters. For example, the deviations appear when we select $n_{max}=m_{max}=2$ at the beam energy of 120 MeV/u, but the deviations disappear when we select $n_{max}=m_{max}=5$. For the results at 50 MeV/u, the reconstruction with $n_{max}=m_{max}=1$ can reproduce the direct fitting results with the deviation less than 14\%. However, none of the combinations of $n_{max}$ and $m_{max}$ can well reproduce the direct fitting results as we found at 120 MeV/u.} %The reason is that the long range interaction plays a role and the overlap region of the system enters the spinodal region during its evolution, and result in the nonmonotonic relationship between $b$ and the averaged values of multiplicity or transverse momentum at central collisions\cite{Lili2018} where the relation of $\bar{N}_i(c_i(N_i))\approx N_i$ does not hold.}

One should note that the multiplicities and total transverse momenta obtained in the ImQMD model are overestimated compared with the experimental data, which are related to the stability of initial nuclei\cite{Xujun16,JPYang2021PRC,JAichelin91PR} and cluster formation mechanism \cite{Nebauer99NPA,YXZhang05PRC,Zbiri07PRC,Danielewicz91NPA,Ono2016} in the QMD type models. It may influence the absolute values of reconstructed $\overline{X}_i$ and $\Sigma_{ij}$, but it will not obviously influence the shape of $P(\mathbf{X})$ and the reconstruction.

\section{Bayesian method for reconstructing impact parameter distribution from two-observables}
\subsection{Sorting centrality with K-means}
Before reconstructing the impact parameter distribution from two observables for selected events or the centrality with Bayes's theorem, we need to find a way to sort the events with $\mathbf{X}=\{M_0, p_{t0}^{tot}\}$ and define the centrality of HICs. 
%For single observale, the experimental centrality $c_X$, which is defined as $c_X=\int_{x}^{\infty}P(X)dX$\cite{Frankland}, has been used for select the events. The range of $x$, such as $x_1\le X\le x_2$ can be found to select the events with the sharp-cut off approach~\cite{Cavata1990}, .
Differently from using a single observable, the simultaneous use of $M_0$ and $p_{t0}^{tot}$ will make difficulties in the determination of the upper and lower boundaries of $\mathbf{X}$. One may artificially define the region in the space of $\mathbf{X}$, for example, a rectangle shape, an elliptic shape, or other shapes, to select the events. This ambiguous criterion requires us to find a rule to sort the centrality of HICs in two-observable space, i.e., $M_0$ and $p_{t0}^{tot}$ space. Compared to the traditional method, the unsupervised machine learning clustering algorithms \cite{MaCQueen1967, Arthur2007}, i.e., the $K$-means clustering method, can automatically classify the events into different classes once the number of classes is given.

The $K$-means clustering method is one of the simplest and commonly used unsupervised machine learning algorithms. It tries to find cluster centers that are representative of certain regions of the data without knowing the label of data points. In this work, the dataset $D=\{\mathbf{X}_i=(M_0,p_{t0}^{tot})_i\}$ with $i=1, \cdots N_{event}$ and $N_{event}$=1,000,000 is generated from the ImQMD model. We classify the dataset into $\mathrm{K}$ clusters, i.e., $C=\{C_1, C_2,\cdots, C_K\}$. $C_i$ represents the sub dataset of the $i$th classification, which is realized by the alternation between two steps: assigning each data point to the closest cluster center where the distance is defined by $d_{ij}=\sqrt{(X_i-X_j)^2}$, and then setting each cluster center as the mean of the data points that are assigned to it. When the assignment of instances to clusters, i.e.,
\begin{equation}
	\label{eq:minE}
	E=\sum_{k=1}^K \sum_{X_i\in C_k}|X_i-\mu_k|^2,
\end{equation}
where $\mu_k$ is the center of cluster $k$, no longer changes, the algorithm will be finished. %More event number will smooth the results in Fig.~\ref{fig:fig3-mpt-snsn}, but do not change the classification dramatically.
%The requirement of using this method is the number of clusters. In our case, it means that we need to specify how many centralities we wish to sort.

There is a question raised for us: why we can use the unsupervised $K$-means clustering method to sort the centrality of HICs? It can be answered from Eq. (\ref{eq:minE}). Suppose the total number of event points in the dataset is $N_0$ and the number of event points in $C_k$ is $N_k$. %In the subdateset of $C_k$ cluster, the events belongs to $C_k$ are represented by $i\in C_k$. 
Based on the previous notification and properties of clusters, the centroid of each cluster is written as
\begin{eqnarray}
	\label{mucm0}
	\mathbf{\mu}_k&=&\frac{1}{N_k} \sum_{\mathbf{i}\in C_k}\mathbf{X}_i.
\end{eqnarray}
As we prove in the Appendix~\ref{centrality},%~\ref{centrality}, 
the centroid of cluster is related to the centrality of selected events as follows,
\begin{eqnarray}
	\mathbf{\mu}_k &\approx&\frac{N_0\mathbf{X}^\ast}{N_k}\sum_{\mathbf{X}\in \Omega(C_k)}P(\mathbf{X})\cdot \Delta s\\\nonumber
	&=&\frac{N_0\mathbf{X}^\ast}{N_k}c(C_k).
\end{eqnarray}
Here, $\Delta s=dM_0dp_{t0}^{tot}$ and $c(C_k)$ is the centrality defined from the event points of cluster $C_k$, i.e., 
\begin{equation}
	c(C_k)=\sum_{\mathbf{X}\in \Omega(C_k)} P(\mathbf{X})\cdot \Delta s, %=\sum_{\mathbf{X}=\mathbf{X}_{low}}^{\mathbf{X}_{up}}P(\mathbf{X}),
\end{equation}
which is similar to the idea of experimental centrality by Abelev \textit{et al.} \cite{Abelev2013,Das18,Frankland21,Yousefnia22}. $\mathbf{X}^*$ is a certain value that satisfies the equality of 
\begin{equation}
\sum_{\mathbf{X}\in \Omega(C_k)} P_{\in C_k}(\mathbf{X})\cdot\mathbf{X}\cdot \Delta s=\mathbf{X}^\ast\sum_{\mathbf{X}\in \Omega(C_k)}P(\mathbf{X})\cdot \Delta s
\end{equation}
Thus, the $K$-means clustering algorithm can be used to sort the centrality.

 %This pseudo-events data can be used to check the validity of the proposed method since it contains the exact values of impact parameters.
\begin{figure}[htbp]
	\centering
	\includegraphics[angle=0,scale=0.37,angle=0]{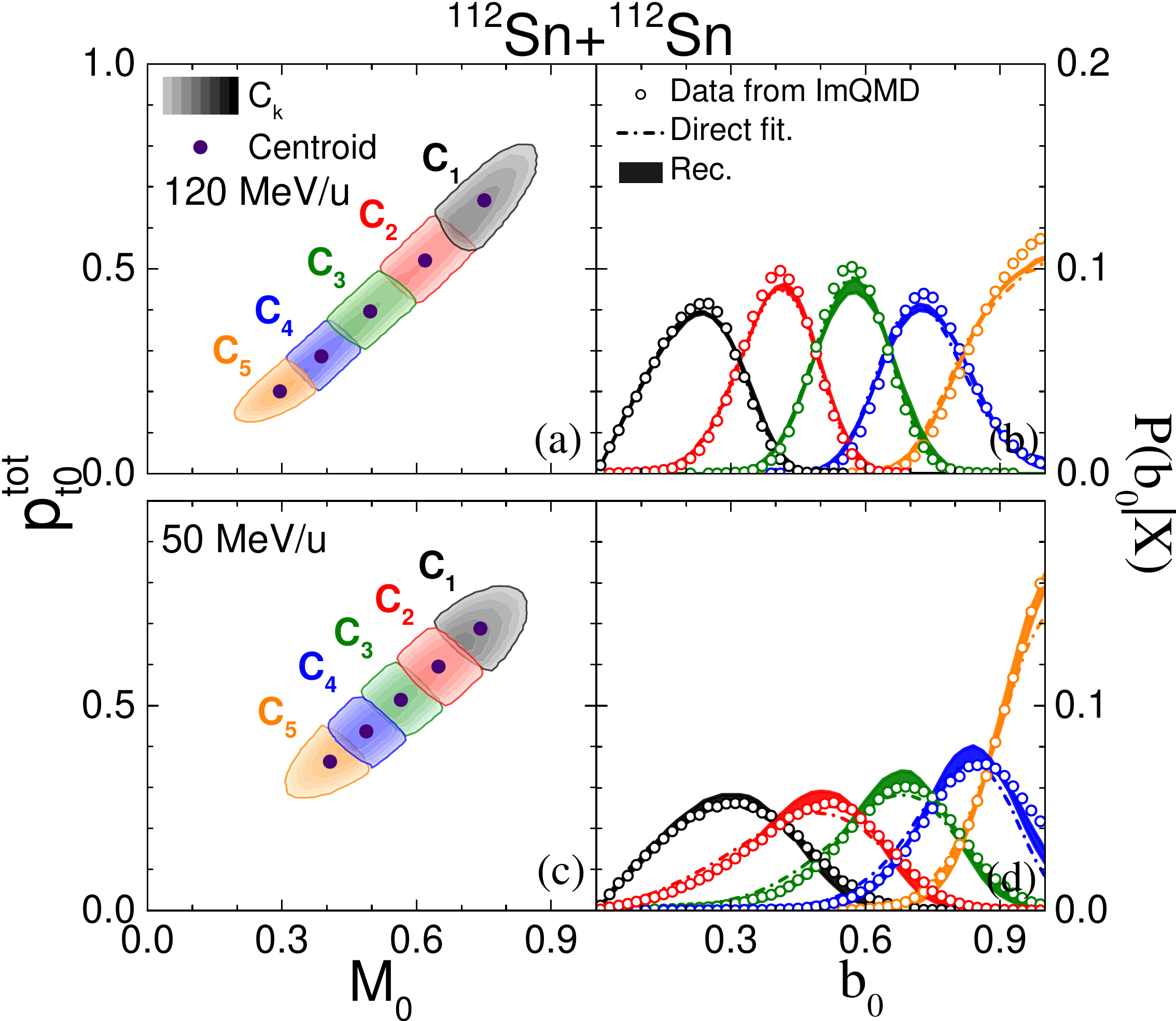}
	\setlength{\abovecaptionskip}{0pt}
	%\vspace{5em}
	\caption{ (a) and (c) Contour plot of $M_0$ vs $p_{t0}^{tot}$ for five clusters. (b) and (d) Reduced impact parameter distributions for five clusters; open circles are the real impact parameter distributions obtained with ImQMD model, and shaded regions are the results inferred with the reconstructing method with different $n_{\mathrm{max}}$ and $m_{\mathrm{max}}$ combinations (dashed lines for direct fitting calculation).}
	\label{fig:fig5-kmeans}
	\setlength{\belowcaptionskip}{0pt}
\end{figure}

%When using unsupervised learning K-means clustering algorithm to determine the centrality of heavy ion collision, each event can be regarded as an event point in the abstract space, and several observables can be taken as the coordinates of the event points, such as $\vec{X_i}=(M,p_t (Z\leqslant 2))$, and each observable represents a dimension of $\vec{X}$.

\subsection{Reconstruction of impact parameter distribution}
Figures \ref{fig:fig5-kmeans}(a) and \ref{fig:fig5-kmeans}(c) show the distributions of event points on the $M_0$ and $p_{t0}^{tot}$ plane for $E_{beam}=$120 and 50 MeV/u, respectively. The events points are sorted into five clusters by the $K$-means clustering algorithm, and are represented by different color regions. The centroids of each cluster are represented by the black solid circles. The overlap between different clusters is less than 10\%, and caused by the algorithm for seeking the minimum of Eq. (\ref{eq:minE}). In Figs. \ref{fig:fig5-kmeans}(b) and \ref{fig:fig5-kmeans}(d), we plot the predicted reduced impact parameter $b_0=b/b_{max}$ distributions by using the Bayesian method, i.e.,
\begin{equation}
	\label{eq:recon}
	P(b|\mathbf{X} \in C_k)=\frac{P(b)\int_{\mathbf{X}\in \Omega(C_k)} P(\mathbf{X}|b)d\mathbf{X}}{\int_{\mathbf{X}\in \Omega(C_k)} P(\mathbf{X}) d\mathbf{X}}. %=\frac{P(b)}{\Delta C_k} \int_{\mathbf{X} \in C_k} P(\mathbf{X}|b) d\mathbf{X}
\end{equation}
Two kinds of $P(\mathbf{X}|b)$ are used. One is direct fitting calculation (dashed lines), and another is the reconstructing method (shaded regions) with different $n_{\mathrm{max}}$ and $m_{\mathrm{max}}$ combinations. The real reduced impact parameter distributions obtained with the ImQMD model (open circles) are used for checking the ability of the two methods. As illustrated in Figs.~\ref{fig:fig5-kmeans}(b) and \ref{fig:fig5-kmeans}(d), the predicted reduced impact parameter distributions from the reconstructing method agree well with the actual impact parameter distributions under the different combinations of $n_{\mathrm{max}}$ and $m_{\mathrm{max}}$, which have $\chi_r^2<$2.

%1) Initializaiton-Gaussian form. In the ImQMD simulations, the events are randomly sampled at given binding energy and radius of the nucleus. Correspondingly, the events are distributed as a Gaussian form in event space. 2) Mean field do not destroy the Gaussian form. As we discussed in Sec.\ref{fluctuation}, the mean-field keeps the . 3) Elastic nucleon-nucleon collisions also do not destroy the Gaussian form.

The reasons why the reconstructing method can reproduce the real impact parameter distribution and is less influenced by the deviations of the covariance matrix can be understood from the following two aspects: one is the validity of Gaussian assumptions on the $P(\mathbf{X}|b)$, and another is the reconstructing method on $P(b|\mathbf{X})$ based on Eq. (\ref{eq:recon}).

The validity of Gaussian assumptions comes from the reaction mechanism, as discussed in Sec. \ref{fluctuation}. The event-by-event fluctuations of final observables with respect to $b$ are dominated by the mechanism of initialization, the mean field potential, and nucleon-nucleon elastic collisions. In the ImQMD simulations, the sampled events are distributed as a Gaussian form in the event space since the initial nuclei in each event are randomly sampled at a given binding energy and radius of the nucleus. The mean field and nucleon-nucleon elastic collisions do not destroy the Gaussian shape of the fluctuations of observables to $b$ at the beam energy we studied. It is a reason why the $\chi_r^2$ is less than 3 in our studies, as shown in Table\ref {tab:chi2}, and are smaller than the $\chi_r^2$ obtained in high energy collisions \cite{Yousefnia22}.%, since the beam energy is close to the transition energy\cite{Zhang06} where the transverse momentum distribute as sphere and $v_2\approx 0$.}

%\textcolor{red}{For the high energy collisions as in Ref.\cite{Yousefnia22}, the inelastic collisions happen and the emitted particles form collective motion at certain rapidity regions due to the strong pressure gradient between the participant and spectators. These behaviors destroy the Gaussian-shaped distribution of final observables, especially away from central collisions.}

\begin{figure}[htbp]
	\centering
	\includegraphics[angle=0,scale=0.37,angle=0]{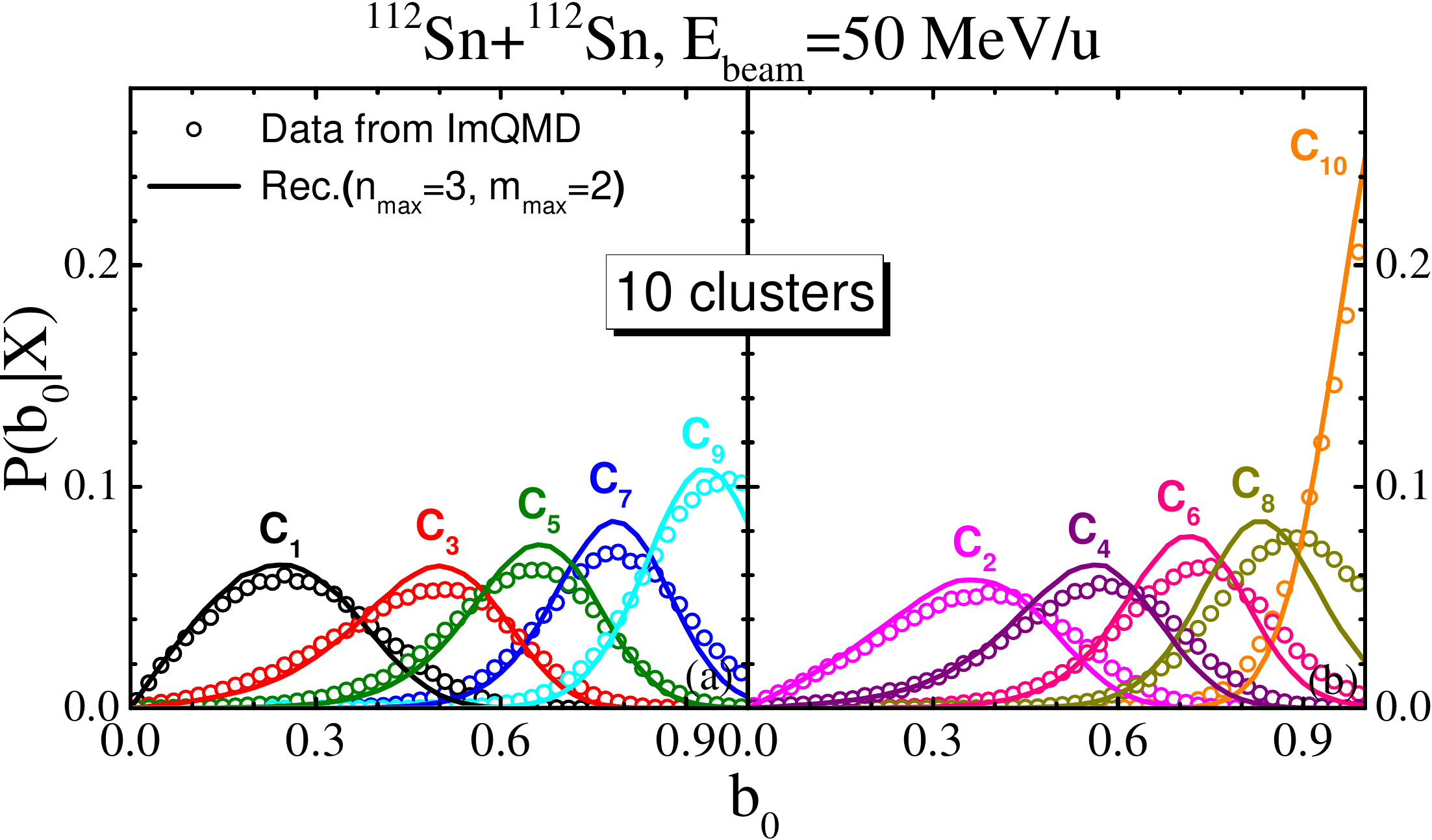}
	\setlength{\abovecaptionskip}{0pt}
	%\vspace{5em}
	\caption{ Same as Fig.~\ref{fig:fig5-kmeans} (d), but for 10 clusters. The lines with different color are the results inferred by the reconstructing method with $n_{\mathrm{max}}=3$ and $m_{\mathrm{max}}=2$.}
	\label{fig:fig9-10cluster}
	\setlength{\belowcaptionskip}{0pt}
\end{figure}

The weak influence of $\Sigma_{ij}$ on the reconstruction of $P(b|\mathbf{X})$, as shown in Fig.\ref{fig:fig5-kmeans}, is related to the range of $\Omega(C_k)$ in the domain of $\mathbf{X}$. The extreme case is to take only 1 cluster by using $K$-means; one can expect that the influence of different values of $\Sigma_{ij}$ completely disappears due to the integration over the full $\mathbf{X}$ space. Quantitatively, in Fig. \ref{fig:fig9-10cluster}, we present the reconstructed $b$ distributions of $^{112}$Sn+$^{112}$Sn at 50 MeV/u for ten clusters. The values $n_{max}=3$ and $m_{max}=2$, which correspond to the largest deviations of $\Sigma_{ij}$ between the reconstructing method and direct fitting calculations, are used to see the effects. The real $b$ distributions for ten clusters are presented as symbols. The left panel is the results from $C_1$, $C_3$, ... to $C_9$, and the right panel is from $C_2$, $C_4$, ... to $C_{10}$. It is clear that the differences between the reconstructed $b$ distribution and real $b$ distribution become larger in the case of ten clusters than that in five clusters [Fig. \ref{fig:fig5-kmeans} (d)].

\section{Summary and discussions}
In summary, we investigate the inherent fluctuation mechanism of intermediate energy heavy ion collisions within the framework of the ImQMD model before studying the reconstruction of the impact parameter distribution from HIC observables. Our calculations show that the inherent fluctuations come from the stochasticity of initialization and nucleon-nucleon scattering in HICs. These inherent fluctuations cause the heavy ion collision observables to fluctuate with $b$, and an accurate determination of the impact parameter is impossible even with as many observables as possible. %Even in the case of perturbative initialization, the distributions of observables are wide enough due to the stochastic nucleon-nucleon collisions.
Thus, the reconstruction of the impact parameter distribution from the selected HIC observables should be done.% instead of to pursue the determination of the exact value of impact parameter.

To model-independently reconstruct the impact parameter distributions for the selected centrality or events for low-intermediate energy HICs, we extend the Bayesian method in which two observables, i.e., multiplicity of charged particles and total transverse momentum of light charged particles, are used simultaneously. A two-dimensional Gaussian-shape fluctuation kernel is adopted, and the parameters of the fluctuation kernel are learned model independently by fitting the pseudoevents data. Since the $b$ dependence of mean values and covariance matrix of experimental data are not known in advance, we also investigate the uncertainties of the extracted $b$ dependence of mean values and covariance matrix. With this form of fluctuation kernel of two observables at a given impact parameter $b$, the impact parameter distributions for selected events can be derived based on Bayes's theorem. For sorting the centrality of heavy ion collisions with multiple observables, we propose to use unsupervised machine learning method, i.e., the $K$-means clustering method, which can automatically select the event sample in the multiobservables space if the class number is given. The validity of using the $K$-means clustering method to sort the centrality of HICs is also proved in the theory. Our calculations show that the reconstructed $b$ distributions agree well with the real $b$ distributions when the number of sorted centrality is around 5 in this energy region. 

Further, the knowledge of the covariance matrix can be used to extract the fluctuations and correlation between the multiplicity of charged particles and the total momentum of light charged particles, which will be useful for learning the fragmentation mechanism.

%Finally, the point one may interest is how about this method applied in the experimental data? To accomplish the task, we need to know the knowledge of experimental filters and observables. The preliminary results based on multiplicity of charged particles has been discussed with \textcolor{red}{experimentalist\cite{Tsang2022}}, and the results seems good. But using this method with multi-observables in low-intermediate energy HIC experiments still need more work to understand the impacts from detector efficiency, angular cover range and so on. 

%For its potential application in the experiments with two-observable, 

%This study will be useful for reconstructing the impact parameter distribuiton from multiplicity of charged particles and transverse momentum of light charged particles.

%investigate the reason , we propose a new model independent method to determine the centrality of heavy ion collision and the distribution of impact parameters. The method is tested by the data given by ImQMD model. Its distinguishing of centrality can well find central, semi-central and peripheral collisions. On the other hand, this method only depends on the spatial distribution of observables, which is conducive to the time classification of transport model calculation in the same way.

\section*{Acknowledgments}
The authors thank the anonymous referee's helpful comments and suggestions. This work was supported by the National Natural Science Foundation of China 
under Grants No. 12275359, No. 11875323, No. 11705163, No. 11790320, No. 11790323, and No. 11961141003, by the National Key R\&D Program of China under Grant No. 2018 YFA0404404, by the Continuous Basic Scientific Research Project (No. WDJC-2019-13, BJ20002501), by funding of the China Institute of Atomic Energy under Grant No. YZ222407001301, and by the Leading Innovation Project of the CNNC under Grants No. LC192209000701 and No. LC202309000201.

\appendix

\section{Relation between the centroid of each cluster and centrality}
\label{centrality}
The centroid of each cluster,
\begin{eqnarray}
	\label{mucm0}
	\mathbf{\mu}_k&=&\frac{1}{N_k} \sum_{\mathbf{i}\in C_k}\mathbf{X}_i,
\end{eqnarray}
can be rewritten as
\begin{eqnarray}
	\label{mucm1}
	\mathbf{\mu}_k&=&\frac{1}{N_k} \sum_{\mathbf{X}\in \Omega(C_k)}\frac{N_{C_k}(\mathbf{X})}{\Delta s}\cdot\mathbf{X}\cdot\Delta s\\\nonumber
	&=&\frac{N_0}{N_k}\sum_{\mathbf{X}\in \Omega(C_k)} P_{\in C_k}(\mathbf{X})\cdot\mathbf{X}\cdot \Delta s
\end{eqnarray}
Here, $N_{C_k}(\mathbf{X})$ is the number of events in cluster $C_k$ in the interval $\Delta s=dM_0dp_{t0}^{tot}$. The values of $\mathbf{X}$ in the $C_k$ cluster appear in the domain of $\Omega$, and their probability density function is $P_{\in C_k}(\mathbf{X})$. In the K-means clustering algorithm, the overlapped event points between different clusters are less than 10\%, and thus, $P_{C_k}(\mathbf{X})\approx P(\mathbf{X})$. Consequently, the centroid of each cluster $\mu_k$ can be approximately described as follows:
\begin{eqnarray}
	\mathbf{\mu}_k &\approx&\frac{N_0\mathbf{X}^\ast}{N_k}\sum_{\mathbf{X}\in \Omega(C_k)}P(\mathbf{X})\cdot \Delta s\\\nonumber
	&=&\frac{N_0\mathbf{X}^\ast}{N_k}c(C_k).
\end{eqnarray}
$c(C_k)$ is the centrality defined from the event points of cluster $C_k$, i.e., 
\begin{equation}
	c(C_k)=\sum_{\mathbf{X}\in \Omega(C_k)} P(\mathbf{X})\cdot \Delta s, %=\sum_{\mathbf{X}=\mathbf{X}_{low}}^{\mathbf{X}_{up}}P(\mathbf{X}),
\end{equation}
The definition is similar to the idea of experimental centrality by Abelev \textit{et al.} \cite{Abelev2013}. $\mathbf{X}^*$ is a certain value that satisfies the equality of 
\begin{equation}
	\sum_{\mathbf{X}\in \Omega(C_k)} P_{\in C_k}(\mathbf{X})\cdot\mathbf{X}\cdot \Delta s=\mathbf{X}^\ast\sum_{\mathbf{X}\in \Omega(C_k)}P(\mathbf{X})\cdot \Delta s
\end{equation} 

\bibliography{References}

\end{document}